\documentclass[aps, prl, twocolumn, superscriptaddress, showpacs, byrevtex, floatfix]{revtex4}
\usepackage{graphicx}

\begin{document}

\title{Ultrafast Spectroscopy of Mid-Infrared Internal Exciton Transitions of \\
Separated Single-Walled Carbon Nanotubes}

\author{Jigang Wang}
\affiliation{Materials Sciences Division, E.O. Lawrence Berkeley National Laboratory, Berkeley, California 94720, USA}
\affiliation{Department of Physics and Astronomy and Ames Laboratory, Iowa State University, Ames, Iowa 50010, USA}
\author{Matt W. Graham}
\affiliation{Department of Chemistry, University of California at Berkeley and Physical Biosciences Division, E. O. Lawrence Berkeley National Laboratory, Berkeley, California 94720, USA}
\author{Yingzhong Ma}
\affiliation{Department of Chemistry, University of California at Berkeley and Physical Biosciences Division, E. O. Lawrence Berkeley National Laboratory, Berkeley, California 94720, USA}
\affiliation{Chemical Sciences Division, Oak Ridge National Laboratory, Oak Ridge, TN 37831, USA}
\author{Graham R. Fleming}
\affiliation{Department of Chemistry, University of California at Berkeley and Physical Biosciences Division, E. O. Lawrence Berkeley National Laboratory, Berkeley, California 94720, USA}
\author{Robert A. Kaindl}
\affiliation{Materials Sciences Division, E.O. Lawrence Berkeley National Laboratory, Berkeley, California 94720, USA}

%\date{November 2, 2009}

\begin{abstract}
We report a femtosecond mid-infrared study of the broadband low-energy response of individually separated (6,5) and (7,5) single-walled carbon nanotubes. Strong photoinduced absorption is observed around 200 meV, whose transition energy, oscillator strength, resonant chirality enhancement and dynamics manifest the observation of quasi-1D intra-excitonic transitions. A model of the nanotube 1$s$-2$p$ cross section agrees well with the signal amplitudes. Our study further reveals saturation of the photoinduced absorption with increasing phase-space filling of the correlated $e$-$h$ pairs.
\end{abstract}
\maketitle

The quasi-1D confinement of photoexcited charges in single-walled carbon nanotubes (SWNTs) gives rise to strongly enhanced Coulomb interactions and large exciton binding energies on the 100 meV energy scale. These amplified electron-hole ({\em e-h}) correlations are a key aspect of nanotube physics \cite{Dre07}. With the availability of individually separated SWNT ensembles, this strong excitonic behavior was confirmed by interband absorption-luminescence maps \cite{Connelletalscience}, two-photon excited luminescence \cite{wangetal2005, Maultzsch2005}, and ultrafast spectroscopy \cite{Mabook2008}. Optical interband probes, however, are limited by symmetry and momentum to detect only a small subset of excitons.

\begin{figure}[!b]
\includegraphics[width=8.3 cm]{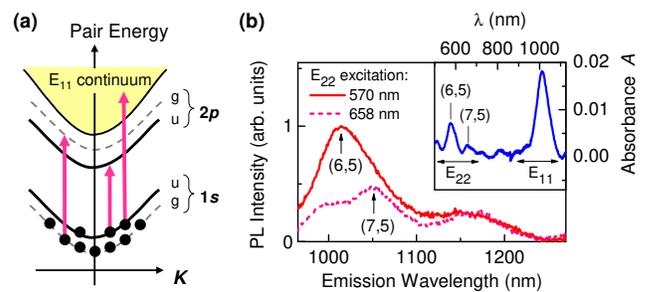}
\caption{(Color online) (a) Two-particle {\em e-h pair} dispersion, illustrating exciton bands and 1s$\rightarrow$2p intra-excitonic transitions (arrows). (b) PL spectra of the sample under resonant $E_{22}$ excitation. Inset: Near-IR absorbance after subtracting background scattering.}
\label{Fig1}
\end{figure}

As illustrated in Fig.~\ref{Fig1}(a) in a two-particle scheme, SWNT excitons are characterized by a center-of-mass momentum {\em K} and by an internal quantum state (designated here as 1{\em s}, 2{\em s}, 2{\em p}, ...) that accounts for the relative charge motion. Each state splits into even (g) and odd (u) parity levels corresponding to different superpositions of the cell-periodic wavefunctions of the underlying graphene lattice \cite{Maultzsch2005}. This entails a series of optically "dark" excitons, including the 1$s$-(g) lowest-energy exciton that lacks coupling in both single- and two-photon interband spectroscopy \cite{wangetal2005, Maultzsch2005}. Splitting into singlet and triplet spin states additionally restricts interband optical coupling \cite{jiangetal2007}. Finally, interband transitions are limited to excitons around K $\approx$ 0 due to momentum conservation.

{\em Intra-excitonic} transitions between low-energy levels of excitons with the same cell-periodic symmetry [arrows, Fig.~\ref{Fig1}(a)] represent a fundamentally different tool, analogous to atomic absorption spectroscopy. In contrast to interband absorption that measures the ability to {\em generate e-h} pairs, intra-excitonic absorption detects existing excitons via transitions from the $1s$ ground state to higher relative-momentum states \cite{Ide08,KaindletalNature03}. Being independent of $K$, it is sensitive to genuine exciton populations across momentum space. As the cell-periodic component of the wavefunction remains unchanged, intra-excitonic absorption is also unrestricted by the exciton ground state symmetry \cite{Ide08}. Applied to individualized SWNTs, intra-excitonic resonances can thus measure both bright and dark excitons and should occur in the mid-infrared (mid-IR) after ultrafast excitation. In contrast to extensive interband nanotube studies \cite{Ma04,ultrafastNIR,Lueretal09, Mur09},  only a few ultrafast intra-band experiments have been carried out which focus largely on nanotube bundles \cite{THz1, BeardetalNL, zhaoetalPRB,luer2009}. THz experiments on photoexcited bundled tubes revealed a non-Drude response attributed to small-gap metallic tubes or inter-tube charge separation \cite{THz1, BeardetalNL}. Mid-IR transient absorption was also observed in bundled nanotubes and assigned to transitions from allowed to dipole-forbidden excitons \cite{zhaoetalPRB,luer2009}.

In this Letter, we report ultrafast optical-pump, mid-IR-probe studies of individually separated (6,5) and (7,5) SWNTs. Transient spectra after photoexcitation evidence strong mid-IR absorption around 200 meV, in accordance with intra-excitonic transitions of strongly-bound $e$-$h$ pairs in semiconducting nanotubes. The absorption cross section of $4\cdot 10^{-15}$ cm$^{2}$ agrees closely with calculations of quasi-1D intra-excitonic 1$s$-2$p$ dipole transitions presented here. The excitation-wavelength dependence and kinetics further underscore the excitonic origin of the mid-IR response, and its intensity dependence scales quantitatively with a model of phase-space filling expected for quasi-1D excitons. This intra-excitonic absorption represents a sensitive tool to probe correlated $e$-$h$ pairs in SWNTs, unhindered by interband dipole or momentum restrictions.

Ultrafast spectroscopy was carried out in transmission using widely tunable femtosecond (fs) pulses in the mid-IR and visible range. Two near-IR optical parametric amplifiers (OPAs) were pumped by a 1-kHz, 28-fs Ti:sapphire amplifier. Resonant and off-resonant interband excitation was achieved using the frequency-doubled OPA or a fraction of the fundamental. The output of the second OPA was difference-frequency mixed in GaSe to generate $\approx$100
fs mid-IR probe pulses tunable from 4-12 $\mu$m \cite{ExpNotes}. We study individually separated Co-Mo-catalyst grown SWNTs of mainly  (6,5) and (7,5) chiralities \cite{BachiloetalJACS} embedded in {50-$\mu$m} thick polyethylene (PE). Importantly, PE ensures transparency throughout our mid-IR probe range, except for a narrow CH-bend vibration at 178~meV. The films were fabricated by drying PE solutions in decalin mixed with micelle-dispersed SWNTs, after transferring SWNTs suspended in H$_2$O with NaDDBS to the PE solution via ultrasound and thermal treatments. In Fig.~\ref{Fig1}(b), the sample's photoluminescence (PL) spectra for resonant (6,5) and (7,5) $E_{22}$ excitation clearly exhibit the distinct $E_{11}$ emission of these individualized SWNT chiralities, with only weak emission from bundled tubes around 1160 nm \cite{TorrensNL2006}. The absorption spectrum [inset, Fig.~\ref{Fig1}(b)] also exhibits the distinct $E_{11}$ and $E_{22}$ absorption peaks.

\begin{figure}[b]
\includegraphics [width=7.2cm]{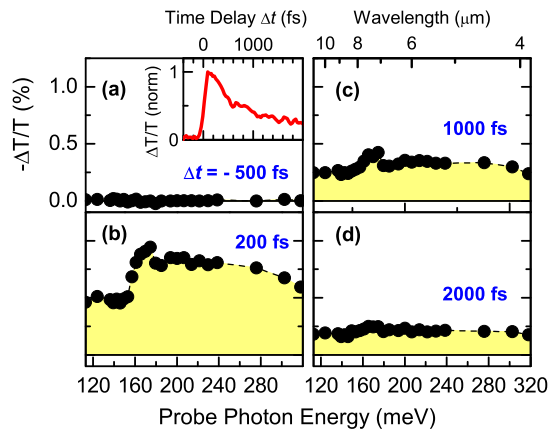}
\caption{(Color online) (a)-(d) Ultrafast spectrally-resolved mid-IR transmission changes after 800 nm excitation for four different delays $\Delta t$ as indicated. Inset: normalized dynamics of the mid-IR transmission probed at 4.4~$\mu$m wavelength.}
\label{Fig2}
\end{figure}

\begin{figure}[!t]
\includegraphics [width = 6.0 cm] {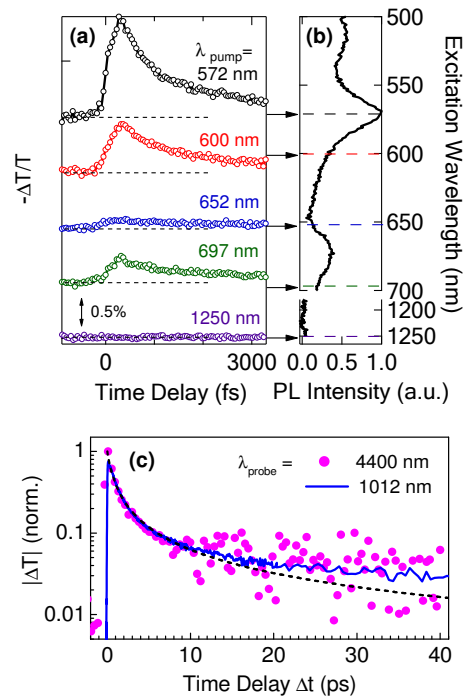}
\caption{(Color online) (a) Pump wavelength dependence on and off-resonant to the (6,5) and (7,5) $E_{22}$ transitions. Traces are offset for clarity, and measured at 4.4~$\mu$m with 260 $\mu$J/cm$^{2}$ excitation fluence. (b) PL-excitation spectrum for fixed $E_{11}$ emission at 1012 nm. (c) Normalized mid-IR dynamics (dots) after 572 nm excitation. Thick line: $E_{11}$ transmission change, scaled to the mid-IR signal at long delays. Dashed line: bimolecular decay $|\Delta T| \propto (1+\gamma~t)^{-1}$ with $\gamma$ = 1.5 ps$^{-1}$.}
\label{Fig3}
\end{figure}

Ultrafast spectrally-resolved mid-IR transmission changes $\Delta T/T$ are shown in Fig.~\ref{Fig2}
for different time delays $\Delta t$ after 800 nm photoexcitation at room temperature.  A strong photo-induced absorption appears within the time resolution after photoexcitation [Fig.~\ref{Fig2}(b), $\Delta t$ = 200 fs] and decays on a ps timescale [Fig.~\ref{Fig2}(c)-(d)].  The transient spectra are characterized by a broadly sloping, asymmetric resonance around 200 meV, with a step-like onset above 160 meV. This mid-IR resonance occurs in the transparent region far below the lowest interband exciton ($E_{11}$ $\simeq$ 1.2 eV) and intersubband transitions ($E_{22}-E_{11}\geq 0.6$eV). The peak energy is close to the (6,5) and (7,5) 1$s$-2$p$ energy splitting in two-photon luminescence studies and calculations \cite{wangetal2005, Maultzsch2005, NoteSplitting}. Thus, we associate this absorption with intra-excitonic transitions between 1$s$ and 2$p$ exciton levels of opposite parity. Both dipole-allowed and optically-dark 1$s$ excitons can fundamentally contribute to this response. The dynamics exhibits a  pulse-width limited rise of the photoinduced mid-IR absorption (inset, Fig.~\ref{Fig2}), which indicates rapid exciton formation.

We should comment on the asymmetric line shape observed in Fig. 2. The observed rapid onset and asymmetry point to a predominantly inhomogeneous broadening, resulting in a large $\simeq$100 meV line width composed of multiple transitions, which matches well with similar spectral features in two-photon PL experiments \cite{wangetal2005, Maultzsch2005}. We attribute the higher-energy tail to intra-excitonic transitions from the 1$s$ into higher-lying n$p$ bound states and into the broad continuum of unbound pairs, consistent with the asymmetric intra-excitonic spectra of quasi-2D $e$-$h$ pairs \cite{KaindletalNature03}. Note that a much narrower peak seems to exist around 170 meV which we assign as an artifact \cite{NoteArtifact}. Low-energy absorption is also observed below 160 meV which can arise e.g. from fluctuations of the dielectric environment around the nanotube and other chiral tube species \cite{JorPRB05b}.

To further substantiate the nature of the response, Fig.~\ref{Fig3}(a) shows the excitation wavelength dependence. Resonant photoexcitation of the (6,5) and (7,5) interband $E_{22}$ transitions, at 572 nm and 697 nm respectively, leads to significant enhancement of the transient mid-IR absorption. The amplitude closely tracks the PL-excitation spectrum [Fig.~\ref{Fig3}(b)], which clearly underscores the tube-specific origin of the transient mid-IR response.  This conclusion  is further supported by the disappearance of the photoinduced signal for excitation below the $E_{11}$ transition [1250 nm, Fig.~\ref{Fig3}(a)]. Hence, the observed photo-induced absorption arises from intra-excitonic transitions of (6,5) and (7,5) SWNTs. The mid-IR dynamics after (6,5) $E_{22}$ excitation is shown in Fig.~\ref{Fig3}(c) on an extended timescale, revealing a strongly non-exponential decay [dots, Fig.~\ref{Fig3}(c)] over several 10 ps. The dynamics closely follows the $E_{11}$ exciton bleaching [solid line, Fig.~\ref{Fig3}(c)], confirming that the mid-IR signals originate from excitations in the $E_{11}$ manifold. Thermal broadening $k_BT\simeq26$~meV at 300 K entails comparable occupation of dark and bright excitons (split by $\simeq10$\,\,meV), which enables this comparison. The decay has a bimolecular shape [dashed line, Fig 3(c)], similar to the fs kinetics of excitons in individualized SWNT suspensions explained by exciton-exciton annihilation \cite{Ma04}, which further underscores the excitonic origin of the mid-IR response.

The mid-IR transmission changes can be used to estimate the absorption cross section $\sigma _{MIR}^{||}$ of the intra-excitonic transition, whose dipole is oriented parallel to the SWNT axis. It is defined as $\sigma _{MIR}^{||}= 3 \, \ln(1-{\Delta T/T})\,/\,n_{exc}$, where $\Delta T$ is the initial transmission change and $n_{exc}$ the photoexcited density. The factor 3 accounts for the random SWNT orientation. Considering (6,5) $E_{22}$ resonant excitation in Fig.~\ref{Fig3}(a) one has ${\triangle T/T} \approx$1.7$\%$ and $n_{exc}=(F/\hbar \omega )\times$ln$(10)A=1.2\times 10^{13}$ cm$^{-2}$, given  $F$ = 260 $\mu$J/cm$^2$ and $A \simeq$ 0.007 [inset, Fig.~\ref{Fig1}(b)]. This yields the experimentally-derived value for the cross section of $\sigma _{MIR}^{||} \simeq 4\times 10^{-15}$ cm$^2$.

\begin{figure}[!t]
\includegraphics [width=6.2 cm] {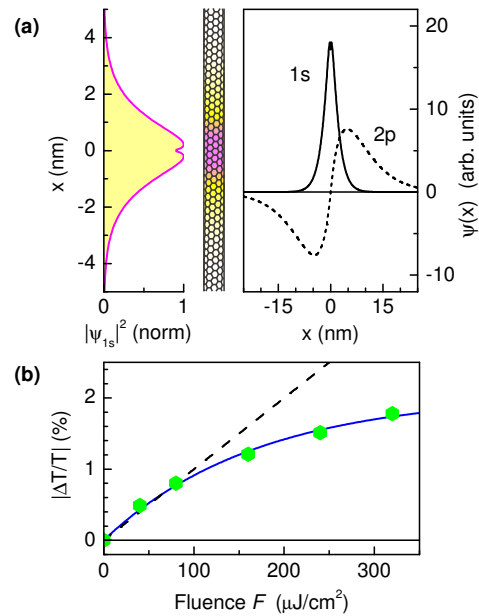}
\caption{(Color online) (a) Squared wavefunction amplitude $\left|\psi_{1s}(x)\right|^{2}$ compared to the (6,5) SWNT scale (left), and bare 1$s$ and 2$p$ wavefunctions (right). (b) pump fluence dependence of the initial mid-IR transmission change (dots) after resonant (6,5) $E_{22}$ excitation. Solid line: model explained in the text, dashed line: linear scaling (guide to the eyes).}
\label{Fig4}
\end{figure}

For comparison, we calculate the intra-excitonic 1$s$-2$p$ cross section based on a model of Wannier-like excitons in SWNTs. The normalized 1$s$ and 2$p$ wavefunctions of Coulomb-bound $e$-$h$ pairs on a cylindrical surface are \cite{corneanetal2007}	

\[
\psi _{1s}(x) = \sqrt{ \frac{8}{(a_B^*)^{3}\alpha_{1s}^3 B_{1s} }} |x|
e^{- \frac{|x|}{a_B^* \alpha_{1s}}}
U(1-\alpha _{1s}, 2, \frac{2|x|}{a_B^*\alpha_{1s}}),
\]
\begin{equation}
\psi _{2p}(x) = \sqrt {2\over (a_{B}^{*})^{3}} \; x \; e^{- \frac{|x|}{a_B^*}},
\end{equation}

\noindent where $x$ measures the distance along the nanotube axis and $a_{B}^{*}=4\pi \epsilon \epsilon _{0}\hbar ^{2}/\mu e^{2}$ is the effective 3D Bohr radius with reduced mass $\mu $ and permittivity $\epsilon $. Furthermore, $U$ is Kummer's confluent hypergeometric function of the second kind and $B_{1s} \equiv 2 \int_{0}^{\infty} y^{2}e^{-y}\left[U(1-\alpha _{1s},2,y)\right]^{2}dy$ is a normalization constant. The binding energies are $E_{1s}=-{\rm Ry}^{*}/(\alpha _{1s})^{2}$ and $E_{2p}= -{\rm Ry}^{*}$, where Ry$^{*}\equiv \hbar ^{2}/(2\mu a_{B}^{*2})$ is the 3D effective Rydberg energy. Also, $\alpha _{1s}$ is a scaling parameter that depends on the SWNT nanotube radius $r_{\rm NT}$ via $\ln(\alpha _{1s})-\Psi (1-\alpha _{1s})-(2\alpha _{1s})^{-1}\equiv \ln(r_{\rm NT})-2\Psi (1)$ where $\Psi$ is the digamma function \cite{corneanetal2007}. For the (6,5) and (7,5) SWNTs studied here, we have $r_{\rm NT}~\simeq~0.4$~nm which entails $\alpha _{1s}$ = 0.33, and $\mu \simeq$  0.067 from interpolated carrier effective masses \cite{Jor05}. The scale and shape of the resulting exciton wavefunctions are shown in Fig.~\ref{Fig4}(a). For this, the permittivity which depends on the local dielectric environment was adjusted to $\epsilon $ = 6 to reproduce the intra-excitonic splitting $E_{2p}-E_{1s}\simeq 0.2\  eV$ from the experiment, which corresponds to a binding energy of 233 meV.

With the above, we obtain the 1$s$-2$p$ intra-excitonic oscillator strength of quasi-1D excitons in SWNTs

\begin{eqnarray}
f_{1s\rightarrow 2p} & \equiv & \frac{128 \, \mu \, a_B^2 \alpha _{1s}^5}{\hbar^2 B_{1s}} (E_{2p} - E_{1s}) \\
\nonumber && \;\;\;\;\; \times \left( \int_{0}^{\infty }s^{3}e^{-s(1+\alpha _{1s})} U(1-\alpha _{1s},2,2s)ds \right) ^{2}
\end{eqnarray}

\noindent For our specific parameters, $f_{1s\rightarrow 2p}$ = 0.41.
Transitions into higher bound $n$p levels ($n >$ 2) were also calculated but are very weak and add less than 15$\% $ in spectral weight. The {\em spectrally-integrated} absorption cross section is then determined as $\sigma _{1s\rightarrow 2p}^{\rm Int} = 2\pi^2 e^2 / (4\pi \epsilon_0 \, \mu c \, n)\times f_{1s\rightarrow 2p}\ =  4.4 \times  10^{-13}$ cm$^2$ meV, where $n$ = 1.5 is the polymer refractive index. Spreading this absorption across $\approx$100 meV results in an estimated 1$s$-2$p$ intra-excitonic cross section of $\sigma _{1s\rightarrow 2p} \simeq 4.4 \times 10^{-15}$ cm$^2$, in very close agreement with our experiment. Full modeling of the asymmetric mid-IR intra-excitonic line shape in Fig.~2 is beyond the scope of the Wannier-exciton model. However, the above illustrates a general consistency between the observed mid-IR signal amplitude and the quasi-1D 1$s$-2$p$ intra-excitonic cross section, motivating more sophisticated theory to calculate bound-bound and bound-continuum spectra with chirality-specific SWNT wave functions.

The transient mid-IR absorption represents a strong oscillator comparable to the interband absorption. In the photoexcited state, this low-energy oscillator strength is derived via transfer from the interband exciton peaks, i.e. from E$_{11}$ bleaching \cite{ultrafastNIR,Lueretal09}.  As plotted in Fig.~\ref{Fig4}(b), with increasing excitation fluence, the mid-IR amplitude $\left|\triangle T/T\right|$ exhibits a distinctly nonlinear behavior. This finding is well described by a saturation model $\triangle T\propto 1-e^{-F/F_{s}}$, shown as the solid line in Fig.~\ref{Fig4}(b) for $F_S$ = 170~$\mu$J$/$cm$^2$. This corresponds to a 1D saturation density $n_{s}=\sigma _{22}^{\rm eff}\times F_{s}/\hbar \omega $, where $\sigma _{22}^{\rm eff}$ is the effective $E_{22}$ absorption cross section per unit nanotube length. The cross section of (6,5) SWNTs was recently found to be  $\sigma _{22}^{||} \simeq$ 85 nm$^2 / \mu$m for light polarized parallel to the nanotube axis, such that  $\sigma _{22}^{\rm eff}=\  1/3\times \sigma _{22}^{||}$ \cite{Berciaudetal2008}. This yields from our experiment a saturation density $n_S = 1.4 \times 10^6$ cm$^{-1}$ corresponding to an average exciton spacing $d_{XX} \approx$ 7 nm. The value is close to the saturation density extrapolated from $E_{11}$ interband bleaching at lower densities \cite{Lueretal09}, while surpassing the saturation of time-averaged PL by more than an order of magnitude \cite{Mur09}. The difference occurs since PL depends on density-dependent decay times that saturate at lower densities, while our study detects the initial pair density. For comparison, we consider phase-space filling (PSF), i.e. the increasing occupation of the constituent Fermion states of the exciton many-particle wavefunction \cite{ Schmittrinketal85}. The PSF density is given by $N_S^{\rm PSF} = L\,/\,[\sum_{k} \psi_k |\psi_k |^2/\psi(x=0)],$ where $\psi _{k}$ are the Fourier coefficients of the exciton wavefunction $\psi(x)$, and $L$ is the normalization length \cite{Lueretal09, Schmittrinketal85,Gre87}. For our quasi-1D 1$s$ exciton wavefunction [Fig.~\ref{Fig4}(a)] this yields $N_{S}^{\rm PSF}=2.5\times 10^{6}$~cm$^{-1}$. Hence, the mid-IR response approaches yet remains somewhat below the limit imposed by phase space filling.

In conclusion, intra-excitonic transitions are both a direct consequence and a measure of {\em e-h} correlations.
Our experiments provide new insights into the chirality-specific femtosecond mid-IR response of electronic excitations in individually separated SWNTs. A photo-induced absorption around 200 meV was observed, manifesting quasi-1D intra-excitonic transitions in close agreement with the calculated 1$s$-2$p$ oscillator strength. The (6,5)/(7,5) chirality-specific enhancement and non-exponential kinetics of the transient mid-IR absorption further underscores its excitonic origin. We believe that the mid-IR probe, extended e.g. into the low-temperature or low-density limit, will provide a versatile spectroscopic tool to investigate bound quasi-1D $e$-$h$ pairs and their internal electronic structure independent of interband symmetry.

%\begin{equation}
%N_S^{\rm PSF} = L \left( \sum_{k} \frac{\psi_k |\psi_k |^2}{\psi(x=0)} \right)^{-1},
%\end{equation}

This work was supported by the Office of Science, U.S. DOE via contract DE-AC02-05CH11231 with initial provision from LDRD.  Manuscript finalization was also supported by Ames Laboratory, DOE contract DE-AC02-07CH11358. SWNTs were characterized at the Molecular Foundry, and prepared at U.C. Berkeley as supported by NSF. M.W.G. thanks NSERC for a graduate fellowship.


\begin{thebibliography}{10}
\bibitem{ExpNotes} The probe was spectrally resolved (10 nm resolution).
\bibitem{NoteSplitting} Differences in this splitting can arise from the dissimilar embedding matrices i.e. dielectric constants.
\bibitem{NoteArtifact} It appears due to a downward step at exactly the 178 meV vibrational resonance in the PE matrix.
\end{thebibliography}
\end{document}